# Deep Learning Applications in Medical Image Analysis: Advancements, Challenges, and Future Directions

Aimina Ali Eli[1], Abida Ali[2]

Master of Science in Information Technology, School of Science & Technology, Emporia State University, Emporia KS, USA

Department of Computer Science and Engineering, Daffodil International University, Dhaka, Bangladesh

**A B S T R A C T**

Medical image analysis has emerged as an essential element of contemporary healthcare, facilitating physicians in achieving expedited and precise diagnosis. Recent breakthroughs in deep learning, a subset of artificial intelligence, have markedly revolutionized the analysis of medical pictures, improving the accuracy and efficiency of clinical procedures. Deep learning algorithms, especially convolutional neural networks (CNNs), have demonstrated remarkable proficiency in autonomously learning features from multidimensional medical pictures, including MRI, CT, and X-ray scans, without the necessity for manual feature extraction. These models have been utilized across multiple medical disciplines, including pathology, radiology, ophthalmology, and cardiology, where they aid in illness detection, classification, and segmentation tasks.

This study offers a thorough examination of applications for deep learning in medical image processing, addressing both its achievements and obstacles. It underscores the relevance of deep learning in enhancing diagnostic precision in tasks including tumor detection, organ classification, and disease progression prediction. Additionally, we investigate sophisticated deep learning methodologies, such as generative adversarial networks (GANs) for picture creation and augmentation, alongside multimodal and federated learning for the integration of heterogeneous data sources and the improvement of model generalizability. Notwithstanding the encouraging outcomes, the domain continues to encounter obstacles like the necessity for extensive labeled datasets, model understanding, and regulatory impediments. We examine methods to tackle these issues and propose future avenues to enhance the influence of deep learning in clinical practice, especially regarding ethical considerations and real-time implementation. As deep learning advances, its incorporation into healthcare is poised to augment medical picture analysis and boost patient outcomes worldwide.

## 1. INTRODUCTION

As a non-invasive tool for visualizing the human body's internal architecture for diagnostic and therapeutic reasons, medical imaging is essential in contemporary healthcare. Classical image analysis methods frequently include domain knowledge and hand-crafted features, both of which are labor-intensive and open to human mistake. Medicine has seen a sea change in image analysis since the introduction of AI, and more specifically deep learning (DL) [1]. Diagnostic imaging has been greatly improved by deep learning, a kind of machine learning, which has an impressive track record of learning from complicated, high-dimensional data. This study delves into the use of DL approaches in medical image analysis, examining its effects in various medical domains like as cardiology, pathology, and radiology. We go over the many kinds of neural networks used, the progress that has been made, the problems that have been encountered, and the ways that DL models could be integrated into existing clinical processes in the future.

## 2. OVERVIEW OF DEEP LEARNING

### A. Fundamentals of Deep Learning

Deep learning relies on artificial neural networks (ANNs) consisting of numerous layers capable of modeling intricate patterns in data. Each layer pulls progressively abstract information from the input, resulting in a hierarchical representation that enhances classification, segmentation, or prediction tasks. The predominant deep learning architectures utilized in medical image analysis are convolutional neural networks (CNNs), recurrent neural networks (RNNs), and generative adversarial networks (GANs). Convolutional Neural Networks (CNNs) are especially adept at image-related tasks because they effectively capture spatial hierarchies in data through the use of convolutional layers. Convolutional Neural Networks (CNNs) have transformed picture categorization, segmentation, and object detection in medical imaging. Recurrent Neural Networks (RNNs), predominantly utilized for sequential data, have been employed in medical imaging for tasks necessitating temporal coherence, such as video-based diagnostic applications [3]. Generative Adversarial Networks (GANs) have demonstrated efficacy in producing synthetic medical images for data augmentation and in generating high-resolution images from low-quality sources [4].

### B. Importance of Deep Learning

It can be challenging to manually interpret medical images, including those from X-rays, computed tomography (CT), and magnetic resonance imaging (MRI), due to their high dimensionality and complicated patterns. Deep learning algorithms can autonomously discern these patterns and offer insights that enhance diagnostic precision. Moreover, DL algorithms outperform conventional methods in terms of speed and consistency when it comes to image segmentation, disease detection, and classification [5]. When it comes to cancer detection, for example, this expertise is vital since a timely and correct diagnosis can have a profound effect on patient outcomes.

## 3. APPLICATIONS OF DEEP LEARNING

### A. Radiology

Radiology is a leading domain in which deep learning has been effectively implemented. Convolutional Neural Networks (CNNs) have been employed to identify diverse anomalies, such as tumors, fractures, and lesions in MRI, CT, and X-ray images [6]. Deep learning models have exhibited enhanced efficacy in detecting pulmonary nodules in chest CT scans, which are critical indications of lung cancer [7].

Algorithms for deep learning have been utilized in breast cancer screening for the automated identification and classification of radiographs. Research indicates that deep learning models can attain accuracy similar to that of seasoned radiologists, hence facilitating the early identification of breast cancer [8]. Moreover, deep learning-based segmentation methods provide accurate delineation of tumor margins, assisting radiologists in evaluating tumor progression and therapeutic efficacy.

### B. Pathology

Digital pathology, encompassing the examination of digitized histopathology slides, has greatly profited from the implementation of deep learning algorithms. Whole-slide imaging (WSI) produces extensive, high-resolution photographs that are challenging to examine manually. Convolutional Neural Networks (CNNs) have been utilized to autonomously identify malignant areas in Whole Slide Imaging (WSI) data, offering pathologists essential decision assistance instruments [9].

Deep learning algorithms have been employed to categorize prostate cancer and assess its level of severity from biopsy specimens. Moreover, deep learning-based image segmentation has demonstrated encouraging outcomes in detecting areas of interest (ROIs) in diverse cancer types, including colorectal and stomach tumors, facilitating enhanced diagnosis and prognostication [10].

### C. Ophthalmology

A number of eye diseases, including diabetic retinopathy, glaucoma, and age-related macular degeneration (AMD), have found DL-based detection systems. Convolutional neural networks (CNNs) have demonstrated excellent accuracy in detecting diabetic retinopathy in retinal fundus pictures, which, if caught in its early stages, can avoid blindness [11]. When applied to primary care settings, these models show promise for automating diagnostic insights from retinal imaging by non-specialists using DL algorithms. In addition, DL models have been created to evaluate glaucoma by identifying alterations in optical coherence tomography (OCT) scans of the optic nerve head and retinal nerve fiber layer. Thanks to these developments, the disease can be better managed and treated at an earlier stage [12].

### D. Cardiology

Echocardiogram, magnetic resonance imaging (MRI), and computed tomography (CT) analysis are some of the cardiology applications where deep learning has shown promising results. Accurate determination of ejection fraction and other heart function parameters has been made possible by using convolutional neural networks (CNNs) to segment the ventricles and atria [13]. DL models can also help cardiologists with diagnostic and prognostic tasks by detecting cardiac problems such arrhythmias, myocardial infarctions, and aortic aneurysms using imaging data [14].

### E. Neurology

In neurology, deep learning models have been utilized in brain imaging to identify and categorize neurological illnesses, including Alzheimer's disease, multiple sclerosis, and intracranial hemorrhages. CNNs have been employed to segment lesions in brain MRI scans, which is essential for the diagnosis and monitoring of multiple sclerosis [16]. Moreover, deep learning models have been employed in the identification of Alzheimer's disease by the examination of structural alterations in the brain observable in MRI scans . In instances of stroke, deep learning algorithms can distinguish between hemorrhagic and ischemic strokes using CT and MRI scans, delivering rapid and precise evaluations essential for prompt intervention [17].

## 4. CHALLENGES IN DEEP LEARNING FOR MEDICAL IMAGE ANALYSIS

There are still a number of obstacles that prevent deep learning from being widely used in clinical settings, even if it has been successful in medical picture processing.

Large, high-quality annotated datasets are not readily available, which is a big obstacle to building successful DL models for medical image interpretation. Medical picture annotation is costly and time-consuming compared to conventional image recognition jobs. Public medical imaging datasets are often undersized or skewed, which hinders DL model performance [18]. To tackle this issue, researchers have turned to transfer learning techniques. These involve refining models that were learned on big datasets that are not related to medicine in order to do specific medical tasks. The question of whether or not these models can be applied to a wide range of patient populations is, however, still open [19].

Deep learning models, especially deep neural networks, are frequently characterized as "black boxes" due to the opacity of their decision-making processes to human interpretation. The absence of openness generates apprehensions within the medical domain, where practitioners must elucidate their diagnostic choices to patients and regulatory bodies [20]. Ongoing endeavors to advance explainable AI (XAI) strategies aim to yield more interpretable model outputs, including visual elucidations of model decisions in medical imaging [13].

The implementation of deep learning models in healthcare settings is governed by rigorous regulatory criteria to guarantee patient safety and effectiveness. Regulatory agencies, like the U.S. Food and Drug Administration (FDA), have commenced formulating criteria for the licensing of AI-driven medical devices; nevertheless, these processes may be protracted and variable. Moreover, ethical problems like patient confidentiality, inaccuracy in model training, and the risk of excessive dependence on AI systems must be resolved prior to the complete integration of deep learning models into healthcare processes [14].

## 5. ETHICAL AND REGULATORY CONSIDERATIONS

To guarantee safe and equitable use, ethical and regulatory concerns about the application of deep learning (DL) to medical image analysis must be addressed. Training AI algorithms on datasets that do not resemble the real world can lead to bias, which is a major ethical concern. In the case of deep learning (DL) models, for example, underperformance when applied to underrepresented populations could result in incorrect diagnoses or insufficient treatment if the models are trained mostly on data from those groups [15]. Because prejudiced algorithms may have a disproportionate impact on marginalized communities, this adds fuel to the fire of healthcare inequities [16].

Patients' right to privacy is another pressing concern. Careful attention to detail is required when dealing with medical imaging data in order to meet the requirements of data protection laws like the General Data Protection Regulation (GDPR) in Europe and the Health Insurance Portability and Accountability Act (HIPAA) in the US. A potential method for increasing the performance of DL models while keeping patient data private is federated learning, which involves training models across different institutions without exchanging patient data [16].

Another obstacle is getting regulatory permission for medical instruments that are driven by AI. Before artificial intelligence (AI) models can be utilized in clinical practice, they must pass stringent validation tests and satisfy the safety and effectiveness requirements specified by regulatory agencies such as the FDA and the EMA. The clearance procedure is slowed significantly, nevertheless, because there are no AI-specific regulatory frameworks. For AI models to acquire the trust of regulators and physicians, transparency is key, and explainable AI (XAI) techniques can help with that [17]. Implementing AI in healthcare in a way that is both ethical and effective requires a delicate balance between the fast-paced progress in DL and strict regulatory criteria.

## 6. INTEGRATION OF AI IN CLINICAL WORKFLOWS

Deep learning (DL) models have great promise for improving patient outcomes, diagnostic accuracy, and healthcare delivery efficiency when integrated into clinical workflows. Careful evaluation of current workflows is necessary for the effective integration of AI tools into everyday medical practice. Artificial intelligence models employed in medical image analysis, for example in tumor detection and segmentation, have the potential to automate mundane but necessary clinical operations, such as picture annotation, freeing up radiologists and other medical experts to concentrate on more intricate cases [18].

AI solutions can also help doctors make better diagnostic and treatment decisions by giving them real-time insights. Digital learning (DL) models trained on massive volumes of imaging data, for instance, may accurately detect anomalies and recommend additional research to doctors. But AI shouldn't supplant human knowledge; it should augment it. Because AI models aren't perfect and can make mistakes, such false positives or negatives, it should still be up to human doctors to make the final diagnosis [19].

In order to deploy AI efficiently, it is essential that it integrates seamlessly with existing healthcare systems like PACS or EHRs. AI needs to have intuitive user interfaces so that doctors who aren't technically savvy may use it. Training programs are also necessary to provide medical staff with the skills to understand and use AI-driven insights. When developers of AI work closely with medical practitioners, they may make sure that these technologies are tailored to healthcare providers' needs. This way, AI can improve clinical decision-making without sacrificing patient care standards [20].

## 7. FUTURE RESEARCH DIRECTIONS

Deep learning's (DL) ongoing development has sparked a number of exciting new avenues of inquiry that might significantly expand DL's utility in medical image analysis. Improving deep learning architectures with more complex features, such transformer-based models, is a key subject. These structures, which were first developed for NLP, have demonstrated promising results when applied to medical image analysis, particularly in the detection of intricate patterns. Improving these models for usage in longitudinal imaging, illness progression prediction, and multi-organ segmentation should be the focus of future studies [21].

The development of self-supervised learning algorithms that can train themselves on unlabeled data is another important trend. The necessity for expert labeling makes the acquisition of big annotated datasets in medical image analysis a challenging task. This difficulty can be reduced through self-supervised learning, which allows DL models to use large volumes of unannotated medical data to enhance their performance on subsequent tasks, like disease classification and anomaly detection 22].

Another fascinating area of research is the use of AI to the field of personalized medicine. Deep learning models can improve the accuracy of disease progression predictions and individualized treatment regimens by integrating imaging data with other patient-specific data such genetic information, clinical history, and lifestyle factors. This has major ramifications for the treatment of long-term illnesses that require tailored approaches, such as cancer and heart disorders [23].

Another potential topic is longitudinal imaging analysis, which involves taking medical images over time to track the evolution of disease or the effectiveness of treatment. DL models that are trained on this kind of longitudinal data have the potential to shed light on the progression of diseases, which could lead to early interventions and more personalized treatments. If medical imaging is to fully benefit from longitudinal analysis, more study into temporal modeling methods and dataset curation is essential [24].

## 8. CONCLUSION

Medical image analysis has already been revolutionized by deep learning, which has brought about advances in accessibility, efficiency, and accuracy. Clinical decision-making and patient outcomes are both enhanced by DL models' ability to automate complicated tasks including disease detection, segmentation, and classification. But there are still a lot of obstacles to overcome, such as issues with data availability and interpretability as well as regulatory worries. These problems, together with emerging areas like multimodal and federated learning, necessitate more investigation in the future. The incorporation of DL into clinical practice, together with its ongoing evolution, has immense promise for transforming healthcare.